\documentclass[letterpaper,aps,prd,,nofootinbib,twocolumn]{revtex4-1}
\usepackage{graphicx,amsmath,amssymb}
\usepackage{url}

\begin{document}

\title{Learning what the Higgs boson is mixed with}

\author{Ryan Killick}
\email{rkillick@connect.carleton.ca}
\author{Kunal Kumar}
\email{kkumar@physics.carleton.ca}
\author{Heather E.~Logan}
\email{logan@physics.carleton.ca}

\affiliation{Ottawa-Carleton Institute for Physics, Carleton University, Ottawa, Ontario K1S 5B6, Canada}

\date{Aug 29, 2013}

\begin{abstract}
The Standard Model Higgs boson may be mixed with another scalar that does not couple to fermions.  The electroweak quantum numbers of such an additional scalar can be determined by measuring the quartic Higgs-Higgs-vector-vector couplings, which contribute---along with the coveted triple Higgs coupling---to double Higgs production in $e^+e^-$ collisions.  We show that simultaneous sensitivity to the quartic Higgs-Higgs-vector-vector coupling and the triple Higgs coupling can be obtained using measurements of the double Higgs production cross section at two different $e^+e^-$ center-of-mass energies.  Kinematic distributions of the two Higgs bosons in the final state could provide additional discriminating power.
\end{abstract}

\maketitle
\section{Introduction}

Last year's discovery at the CERN Large Hadron Collider (LHC) of a new particle $h$~\cite{:2012gk} consistent with the Higgs boson of the Standard Model (SM) marks the start of a long-term experimental program of measurements of Higgs properties.  Through this program we hope to learn the nature of the particle itself and understand the underlying physics responsible for the breaking of the electroweak symmetry.

Many extensions of the SM contain one or more additional scalar particles that can mix with the SM Higgs boson.  Such scenarios can be tested experimentally through measurements of the couplings of the discovered Higgs boson, which will be modified in general from their expected SM values, as well as through direct searches for the additional scalar particles.  Measurements of the couplings of such a mixed Higgs provide some information about the degree of mixing between the SM Higgs and the additional scalar and the additional scalar's contribution to electroweak symmetry breaking.  If the additional scalar couples to charged fermions (this is possible only for scalars that originate from an SU(2) doublet), coupling measurements will also shed light on the fermion coupling pattern.

Information about the electroweak quantum numbers of the additional scalar, however, cannot generally be obtained from measurements of the three-point couplings of the discovered Higgs boson to pairs of fermions or gauge bosons.  This is easiest to see in the case that the additional scalar does not contribute to electroweak symmetry breaking (we also assume that it does not couple to fermions); in this case, the couplings of the discovered Higgs boson to a pair of SM fermions or gauge bosons are all modified by a common multiplicative factor that depends on the mixing angle between the SM Higgs and the additional scalar.  Measurements of these couplings provide no information about the electroweak quantum numbers of the additional scalar.  Even if the additional scalar does contribute to electroweak symmetry breaking, its electroweak quantum numbers cannot generally be disentangled from mixing effects in the three-point couplings of the discovered Higgs boson.

In this paper we propose a strategy to determine the electroweak quantum numbers of the additional scalar by measuring the four-point coupling of a pair of $W$ or $Z$ bosons to a pair of the discovered Higgs bosons.  This coupling depends on the weak isospin and hypercharge of the additional scalar and can be accessed experimentally in electroweak-initiated double Higgs production.  We consider the processes $e^+e^- \to Zhh$ and $e^+e^- \to \nu \bar \nu hh$ at the proposed International Linear Collider (ILC)~\cite{DBD}.
These processes have previously been studied as a way to measure the Higgs self-coupling at the ILC~\cite{Djouadi:1999gv,Castanier:2001sf,Battaglia:2001nn,Fujii-Princeton}.  We show that the four-point $hhVV$ coupling and the Higgs self-coupling can be simultaneously extracted using measurements of the double Higgs production cross section at two different $e^+e^-$ collision energies.  We also suggest a more sophisticated extraction strategy using the dependence of the two Higgs invariant mass distribution on the two couplings of interest.

This paper is organized as follows.  In Sec.~\ref{sec:couplings} we define a formalism and present the $hhVV$ couplings for three benchmark models.  Section~\ref{sec:measuring} contains our main results.  We conclude in Sec.~\ref{sec:conclusions}.  In the appendices we discuss the situation in which the additional scalar contributes to electroweak symmetry breaking and comment on the experimental and theoretical constraints on our chosen benchmark points.

\section{Couplings of a mixed Higgs boson}
\label{sec:couplings}

The couplings of a Higgs-like scalar field $h$ to SM particles can be parameterized by an effective Lagrangian,
\begin{eqnarray}
	\mathcal{L} &\supset& 
	k_V M_V^2 V_{\mu}^* V^{\mu} \left[ 1 + a_V \frac{2h}{v_{\rm SM}} + b_V \frac{h^2}{v_{\rm SM}^2} \right]  
	- m_f \bar f f \left[ 1 + c_f \frac{h}{v_{\rm SM}} \right] \nonumber \\
	&& - \frac{1}{2} M_h^2 h^2 \left[1 + d_3 \frac{h}{v_{\rm SM}} + d_4 \frac{h^2}{4v_{\rm SM}^2} \right],
	\label{eq:efflagrangian}
\end{eqnarray}
where $v_{\rm SM} = 246$~GeV is the SM Higgs vacuum expectation value (vev), $V = W$ or $Z$, and $k_W = 1$, $k_Z = 1/2$ accounts for the extra symmetry factor in the term involving $Z_{\mu} Z^{\mu}$.  In the SM, the scaling parameters $a_i,b_i,c_i,d_i$ are all equal to 1.  In general one can define a different scaling parameter $c_f$ for each fermion.  Models in which the custodial SU(2) symmetry is violated can have $(a_W,b_W) \neq (a_Z,b_Z)$.

\subsection{SM Higgs mixed with an additional scalar}

We consider the case in which the SM Higgs boson $\phi$ mixes with the real neutral state $\chi$ of a general electroweak multiplet $X$.  The discovered Higgs particle $h$ is then a linear combination of $\phi$ and $\chi$, 
\begin{equation}
	h = \phi \cos\theta - \chi \sin\theta.
\end{equation}

This mixing generically modifies the couplings of $h$ to gauge boson and fermion pairs.  The most difficult scenario to probe experimentally is that in which $X$ does not couple to fermions\footnote{The coupling of a scalar electroweak multiplet $X$ to fermions is forbidden by gauge invariance unless $X$ is an isospin doublet.  (We do not consider the lepton number violating couplings that can be written down involving a scalar triplet~\cite{seesaw}.)  In the latter case, couplings of $X$ to fermions can be forbidden using a discrete symmetry, yielding a Type-I two Higgs doublet model~\cite{Georgi:1978wr,Haber:1978jt}.  Such a symmetry allows the model to avoid stringent experimental constraints on flavor-changing neutral currents~\cite{Glashow:1976nt}.} and does not acquire a vev.  The single-$h$ couplings to gauge bosons and fermions are then given by
\begin{equation}
	a_W = a_Z \equiv a = \cos\theta, \qquad
	c_f \equiv c = \cos\theta.
\end{equation}
In particular, measurements of these couplings tell us nothing about the quantum numbers of $X$.

When $X$ acquires a vev, it contributes to the masses of the $W$ and $Z$ bosons and reduces the vev that can be carried by the doublet.  This drives up the Yukawa couplings of the doublet that are required in order to obtain the observed fermion masses.  It also gives rise to a coupling of $\chi$ to $WW$ and $ZZ$ proportional to the vev of $X$.  In this case, $a \neq c$; in particular the $hVV$ and $h f \bar f$ couplings become (see Appendix~\ref{app:vev} for details)
\begin{equation}
	a_V = \cos\theta \sin\beta - \sqrt{b_V^{\chi}} \sin\theta \cos\beta, \qquad
	c = \frac{\cos\theta}{\sin\beta},
\end{equation}
where $\sin\beta = v_{\phi}/v_{\rm SM}$ is the doublet vev in units of the SM Higgs vev $v_{\rm SM}$, and we define $b_V^{\chi}$ as 
\begin{equation}
	b_W^{\chi} = 2 \left[ T (T+1) - \frac{Y^2}{4} \right], \qquad
	b_Z^{\chi} = Y^2.
	\label{eq:bchidef}
\end{equation}
Here $T$ and $Y$ are the weak isospin and hypercharge of the multiplet $X$ and we have used $Q = T^3 + Y/2 = 0$ to simplify the expressions for the real neutral state $\chi$.
In this notation, the couplings of the SM Higgs $\phi$ (with $T = 1/2$, $Y = 1$) are $b_W^{\phi} = b_Z^{\phi} = 1$.

Note in particular that the two couplings, $a_V$ and $c$, depend on three a priori unknown parameters, $b_V^{\chi}$, $\cos\theta$, and $\sin\beta$.  The parameter $b_V^{\chi}$ thus cannot be extracted from measurements of three-point Higgs couplings even in the case that the additional multiplet $X$ carries a vev.

\subsection{$hhVV$ coupling}

The $hhVV$ coupling receives contributions from both the $\phi\phi VV$ and $\chi\chi VV$ couplings.  These couplings are independent of the vevs of $\phi$ and $X$.  The relevant Feynman rules in the electroweak basis are 
\begin{equation}
	\chi\chi W^+_{\mu}W^-_{\nu}: \ i \frac{g^2}{2} b^{\chi}_W g_{\mu\nu}, \quad
	\chi\chi Z_{\mu} Z_{\nu}: \ i \frac{g^2}{2 c_W^2} b^{\chi}_Z g_{\mu\nu},
\end{equation}
where $b_V^{\chi}$ are defined in Eq.~(\ref{eq:bchidef}).

After mixing, the $hhVV$ coupling scaling factors $b_V$ become
\begin{equation}
	b_V = \cos^2 \theta + b_V^{\chi} \sin^2 \theta.
	\label{eq:bV}
\end{equation}
This coupling depends only on the mixing angle $\theta$ and the electroweak quantum numbers of $X$.  When $X$ does not carry a vev, measurements of $a$ and/or $c$ fix the mixing angle and a measurement of $b_V$ can then be used to determine these quantum numbers.  When $X$ carries a vev, measurements of $a_V$, $c$, and $b_V$ provide enough information to unambiguously determine the electroweak quantum numbers of $X$.

Note that while the $hhWW$ and $hhZZ$ couplings in general have different scaling factors, models that preserve custodial SU(2) symmetry will have $b_W = b_Z$.  We consider such models in what follows.

\subsection{Three benchmark models}

\noindent
\underline{SM mixed with a singlet:} In this case the discovered Higgs boson $h$ is a mixture of the SM doublet Higgs field $\phi$ and a real singlet scalar $s$~\cite{Barger:2007im},
\begin{equation}
	h = \phi \cos\theta - s \sin\theta.
\end{equation}
The singlet does not couple to SM fermions or gauge bosons, so $a = c = \cos\theta$.
The couplings $b_V$ are then given by
\begin{equation}
	b_W = b_Z \equiv b = \cos^2\theta = a^2.
\end{equation}
This relationship between $b$ and $a$ holds regardless of whether the singlet carries a vev.

\vspace{0.5cm}

\noindent
\underline{SM mixed with an additional doublet:}  
In a model containing two (or more) Higgs doublets, the $hhVV$ couplings are given by 
\begin{equation}
	b_W = b_Z \equiv b = 1,
\end{equation}
regardless of the vevs of the doublets and their couplings to fermions.\footnote{The most common formulation of two Higgs doublet models implements a softly-broken $Z_2$ symmetry in the Higgs potential, which is used to enforce the fermion coupling structure~\cite{Glashow:1976nt}.  This symmetry results in $a = c = 1$ when the vev of the second doublet is set to zero.  When the second doublet carries a vev, $a \neq c$ in general, allowing the two-doublet model to be distinguished from the SM mixed with a singlet.  If no $Z_2$ symmetry is imposed, additional quartic Higgs couplings appear in the potential that allow $a = c \neq 1$ even when the vev of the second doublet is zero~\cite{Enberg:2013ara}.  In this case a theory of flavor must be invoked to explain the absence of flavor-changing neutral Higgs couplings.}

\vspace{0.5cm}

\noindent
\underline{A custodial SU(2)-preserving model with triplets:}
The Georgi-Machacek (GM) model~\cite{Georgi:1985nv,Chanowitz:1985ug,Gunion:1989ci} contains the SM Higgs doublet together with two Higgs triplets---a complex triplet with hypercharge $Y = 2$ in our conventions and a real triplet with hypercharge $Y=0$---arranged in such a way as to preserve custodial SU(2) symmetry.\footnote{Custodial symmetry is also preserved by the SM Higgs doublet mixed with a scalar septet ($T=3$, $Y=4$)~\cite{Hisano:2013sn}, which yields a staggering $b_W^{\chi} = b_Z^{\chi} = 16$.  For the benchmark value $a = 0.9$ that we will consider and neglecting the septet vev, this leads to $b = 3.85$, which will be well separated experimentally from the singlet, doublet, and triplet models (see Fig.~\ref{fig:chisq}).  Even for a very small mixing of $a = 0.99$, the septet model still yields a sizable $b = 1.30$.}  
The model contains two singlets of custodial SU(2):
\begin{equation}
	H_1^0 = \phi, \qquad \qquad 
	H_1^{0 \prime} = \sqrt{\frac{2}{3}} \chi^{0,r} + \frac{1}{\sqrt{3}} \xi^0,
\end{equation}
where $\chi^{0,r}$ is the real neutral component of the complex triplet and $\xi^0$ is the neutral component of the real triplet.

We assume that the observed state $h$ is a custodial SU(2)-preserving mixture of $H_1^0$ and $H_1^{0 \prime}$, 
\begin{equation}
	h = H_1^0 \cos\theta - H_1^{0 \prime} \sin\theta.
\end{equation}
The couplings $b_V$ of $h$ are given by,\footnote{If only one of the two triplets were present, $b_W$ and $b_Z$ would not be equal.  In particular, if $X$ is a real triplet ($T = 1$, $Y = 0$), $b_W = \cos^2\theta + 4 \sin^2\theta$ and $b_Z = \cos^2\theta$; similarly, if $X$ is a complex triplet ($T = 1$, $Y = 2$), $b_W = \cos^2\theta + 2 \sin^2\theta$ and $b_Z = \cos^2\theta + 4 \sin^2\theta$.  Phenomenology of the real triplet has recently been studied in Ref.~\cite{FileviezPerez:2008bj}.}
\begin{equation}
	b_W = b_Z \equiv b = \cos^2\theta + \frac{8}{3} \sin^2\theta.
\end{equation}
If we assume that the vevs of the triplets are zero (so that $c_H = 1$ in the notation of Ref.~\cite{HHG}), $b$ can be expressed in terms of $a$ according to,
\begin{equation}
	b_W = b_Z = a^2 + \frac{8}{3} (1 - a^2).
	\label{eq:GMb}
\end{equation}
This assumption is problematic in that the mixing angle $\theta$ goes to zero in the limit that the triplet vevs vanish~\cite{GMpaper}.  Nevertheless, Eq.~(\ref{eq:GMb}) holds approximately when the triplet vev is sufficiently small.  A prescription for determining $b$ in terms of $a$ and $c$ in the case of nonzero triplet vev is given in Appendix~\ref{app:vev}.

\section{Measuring the $hhVV$ coupling}
\label{sec:measuring}

We consider the scenario in which measurements of Higgs couplings at the LHC and an early-stage 250~GeV ILC have revealed deviations in the Higgs couplings consistent with mixing between the SM Higgs and a scalar that does not couple singly to SM gauge bosons or fermions.  For concreteness we take $a = c = 0.9$.  In this case, all single Higgs production rates at LHC and ILC are reduced to $a^2 = c^2 = 0.81$ times their SM values, while all Higgs branching ratios are the same as predicted in the SM.  The most precise direct measurement of $a$ at this stage will come from the 250~GeV ILC measurement of the inclusive $e^+e^- \to Zh$ cross section, yielding $\Delta a/a = 1.3$\% after 250~fb$^{-1}$~\cite{DBD}, i.e., a deviation from $a=1$ with significance 7.8$\sigma$.

\subsection{Double Higgs production at ILC}

\begin{figure}
\resizebox{0.16\textwidth}{!}{\includegraphics{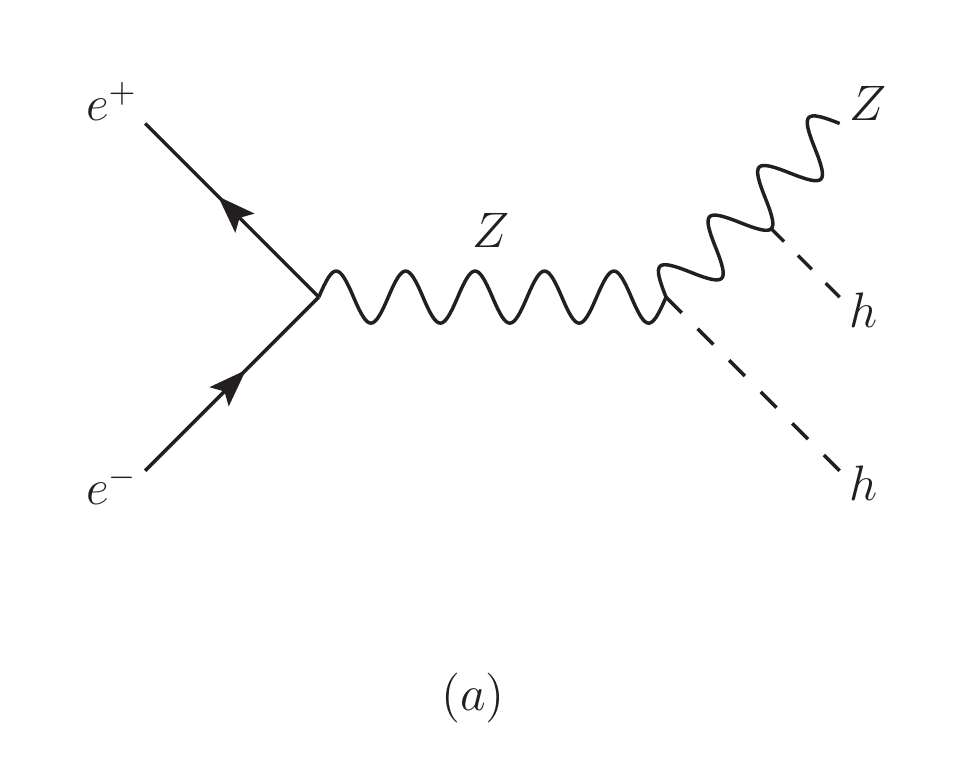}
}\resizebox{0.16\textwidth}{!}{\includegraphics{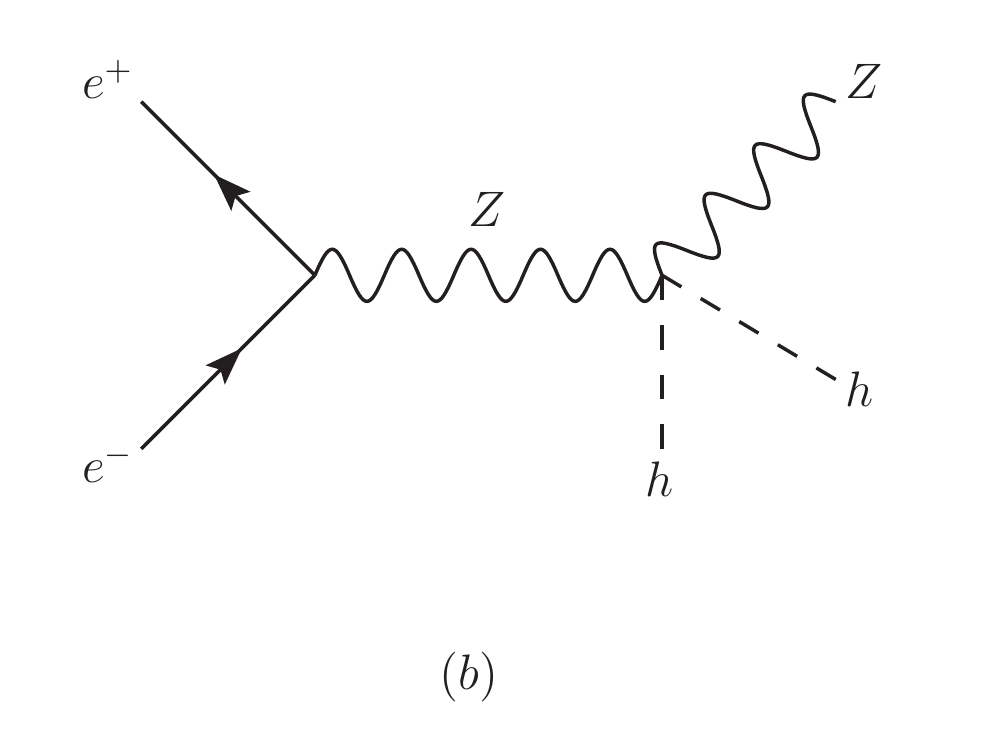}
}\resizebox{0.16\textwidth}{!}{\includegraphics{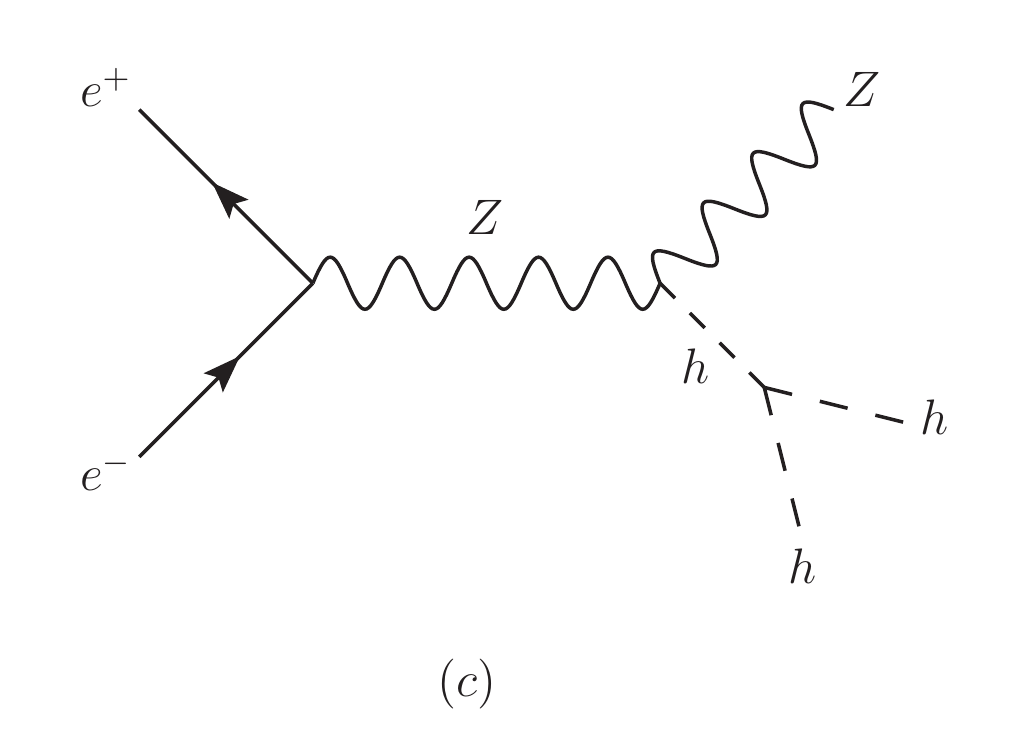}}
\caption{Feynman diagrams for $e^+e^- \to Zhh$.  We include the crossed version of diagram (a).}
\label{fig:eeZhh}
\end{figure}

\begin{figure}
\resizebox{0.13\textwidth}{!}{\includegraphics{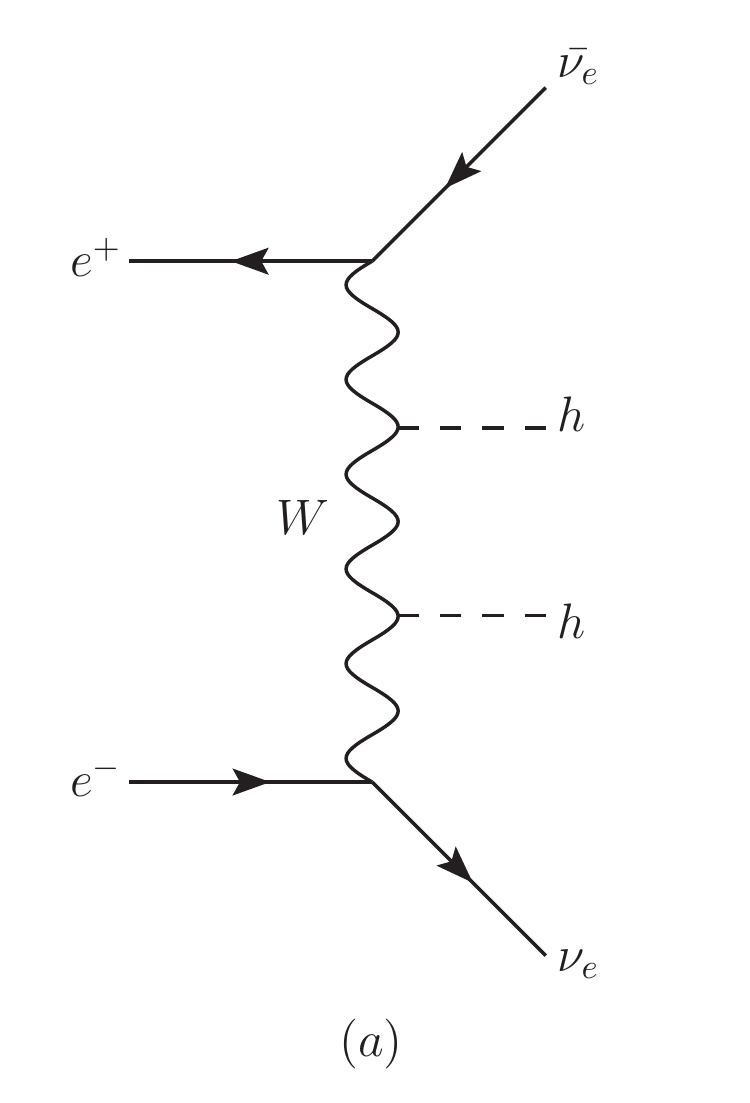}
}\resizebox{0.16\textwidth}{!}{\includegraphics{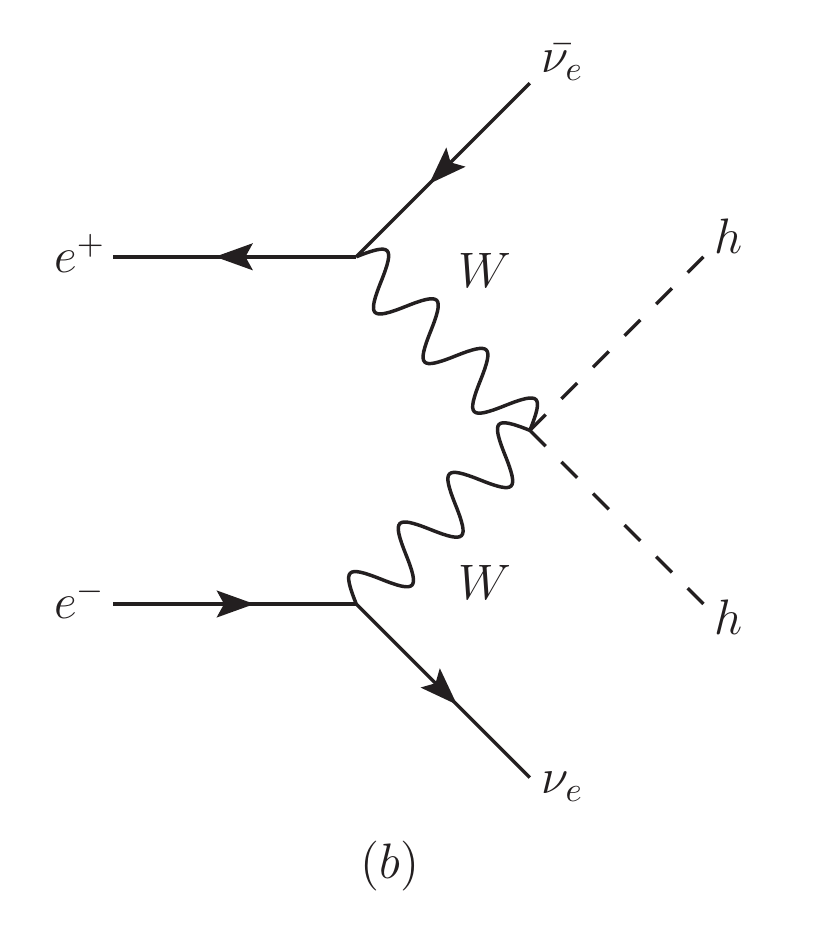}
}\resizebox{0.16\textwidth}{!}{\includegraphics{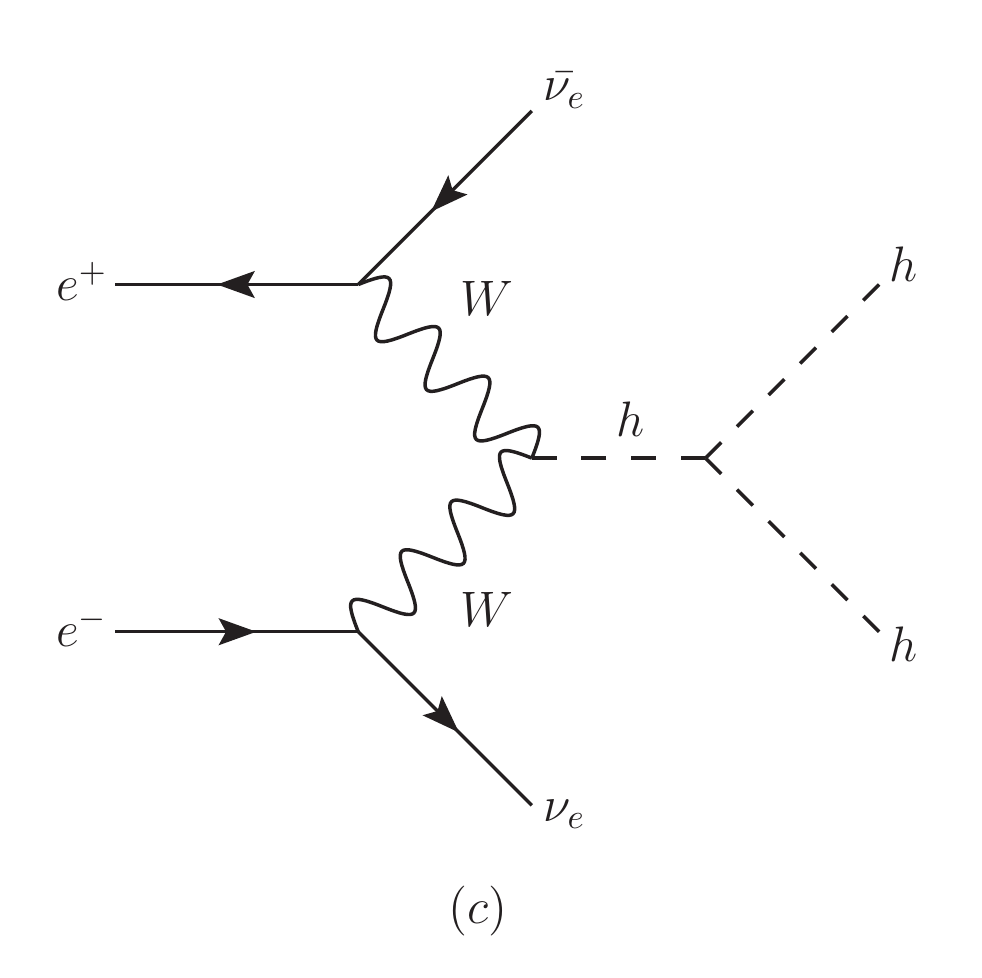}}
\caption{Feynman diagrams for  $e^+e^- \to \nu \bar \nu hh$ via $W$ boson fusion.  We include the crossed version of diagram (a).}
\label{fig:eeWBFhh}
\end{figure}

Double Higgs production in $e^+e^-$ collisions proceeds through the processes $e^+e^- \to Zhh$ and $e^+e^- \to \nu_e \bar \nu_e hh$ via $W$ boson fusion (WBF).  Feynman diagrams are shown in Figs.~\ref{fig:eeZhh} and~\ref{fig:eeWBFhh}.  The cross section for double Higgs production via $Z$ boson fusion, $e^+e^- \to e^+ e^- hh$, is much smaller than that from WBF and we neglect it here.

We computed the double Higgs cross sections in the effective theory described by Eq.~(\ref{eq:efflagrangian}) using the public package {\tt CalcHEP}~\cite{CalcHEP} and checked our results using {\tt MadGraph}~\cite{MadGraph}.  In all cases we set $a = 0.9$ and $d_3 \equiv d = 1$.  The resulting values of $b$ in our three benchmark models are given in Table~\ref{tab:bvalues}, along with the double Higgs production cross sections.  We consider $e^+e^- \to Zhh$ at 500~GeV and 1000~GeV center-of-mass energies as well as WBF at 1000~GeV center-of-mass energy.

\begin{table}
\begin{tabular}{ccccc}
\hline\hline 
Model & $b$ & $\sigma^{500}(Zhh)$ & $\sigma^{1000}(Zhh)$ & $\sigma^{1000}$(WBF) \\
\hline
Singlet & 0.81 & 0.11 fb  &  0.082 fb & 0.041 fb \\
Doublet & 1 & 0.14 fb  &   0.11 fb&  0.027 fb \\
GM & 1.32 & 0.19 fb& 0.18 fb &  0.090 fb\\
\hline
SM & 1 &  0.16 fb &  0.12 fb & 0.071 fb \\
\hline\hline
\end{tabular}
\caption{Values of $b$ and unpolarized signal cross sections (computed using {\tt CalcHEP}~\cite{CalcHEP}) for the three benchmark models with $a = 0.9$.  The SM cross sections are shown for comparison.  Cross sections do not include any $Z$ or $h$ branching ratios.  In all cases we assume $d = 1$. }
\label{tab:bvalues}
\end{table}

Our effective theory does not include the contributions from $t$- and $u$-channel exchange of the custodial SU(2) triplet states $H^{\pm}, A^0$ present in the doublet and GM models, nor does it include processes involving single production of the heavier custodial singlet $H^0$ (decaying to $hh$) that is present in all three of our benchmarks.  For simplicity, we assume that these extra states are heavy enough that their on-shell production is kinematically forbidden at the $e^+e^-$ collision energies that we use in our analysis.  This requires $M_{H^0} \gtrsim 910$~GeV (from $e^+e^- \to ZH^0$ at 1~TeV; WBF production of $H^0$ is severely kinematically suppressed near threshold), $M_{A^0} \gtrsim 875$~GeV (from $e^+e^- \to h A^0$), and $M_{H^{\pm}} \gtrsim 500$~GeV (from $e^+e^- \to H^+ H^-$).  We discuss the viability of this assumption in our specific model benchmarks in Appendix~\ref{app:unitarity}.

We checked using an explicit calculation in the doublet model that including the diagrams involving $H^{\pm}$ and $A^0$ with masses above these thresholds does not significantly change the double Higgs production cross sections of interest.\footnote{For example, a benchmark point with $M_{H^+} = 660$~GeV and $M_{A^0} = 880$~GeV increases $\sigma^{500}(Zhh)$ by less than 1\%, $\sigma^{1000}$(WBF) by about 1.5\%, and $\sigma^{1000}(Zhh)$ by about 7\%.  Because $Zhh$ production contributes only about 10\% of the $\nu\nu hh$ signal rate at 1~TeV after the selection cuts of Ref.~\cite{Fujii-Princeton}, the change in $\sigma^{1000}(Zhh)$ increases the total signal rate by less than 1\%.}

\subsection{Extracting $b$ and $d$ from event rates}

The dependence of the double Higgs production cross sections on $b$ and $d$ varies with center-of-mass energy and with the process considered.  Therefore, measurements of the double Higgs production cross section at two different ILC center-of-mass energies can be used to fit for the parameters $b$ and $d$, given a fixed value of $a$ (assumed to be measured in $e^+e^- \to Zh$).  We perform such a fit using preliminary double Higgs production cross section uncertainties from the ILC Large Detector (ILD) study for the ILC Technical Design Report~\cite{Fujii-Princeton}.    
At 500~GeV the process of interest is $e^+e^- \to Zhh$, with $Z \to e \bar e$, $\mu \bar \mu$, $\nu \bar\nu$, and $q \bar q$.  At 1~TeV the process of interest is $e^+e^- \to \nu \bar \nu hh$, including contributions from WBF and $Z(\to \nu \bar \nu)hh$.

Because the signal rates for our benchmark points are different than the signal rates in the SM, we rescale the statistical uncertainties found for the SM in Ref.~\cite{Fujii-Princeton} based on the number of signal events at our benchmark points.  We also take into account the different selection efficiencies for the $Zhh$ and WBF processes at 1~TeV by scaling our computed $Zhh$ cross section to obtain the same relative efficiency quoted in Ref.~\cite{Fujii-Princeton}.  

The resulting rescaled uncertainties are summarized in Table~\ref{tab:expt-scaled}.  This rescaling increases the fractional uncertainties for the singlet and doublet models and decreases them for the GM model.  We do not attempt to account for the effect of the kinematic cuts on the relative contributions of the diagrams in Figs.~\ref{fig:eeZhh} and~\ref{fig:eeWBFhh} to the total cross sections.   

\begin{table}
\begin{tabular}{ccccc}
\hline\hline 
Model & $b$ & $\Delta \sigma/\sigma (Zhh, 500~{\rm GeV})$ & $\Delta \sigma/\sigma (\nu \nu hh, 1~{\rm TeV})$ \\
\hline
Singlet & 0.81 & 38\%  &  32\% \\
Doublet & 1 & 32\%  &    42\% \\
GM & 1.32 & 24\%  &  18\% \\
\hline
SM & 1 &  27\% &   23\% \\
\hline\hline
\end{tabular}
\caption{Expected experimental uncertainties on the double Higgs production cross sections at the 500~GeV and 1~TeV ILC. The SM values have been taken from Ref.~\cite{Fujii-Princeton}. The uncertainties for the other models have been scaled to account for the change in cross section produced by the different $a$ and $b$ values. 
An integrated luminosity of 2~ab$^{-1}$ is assumed in each case.  The electron and positron beam polarizations have been taken as $P(e^-,e^+) = (-0.8,+0.3)$ at 500~GeV and $(-0.8,+0.2)$ at 1~TeV.}
\label{tab:expt-scaled}
\end{table}

\begin{figure}
\resizebox{0.5\textwidth}{!}{\includegraphics{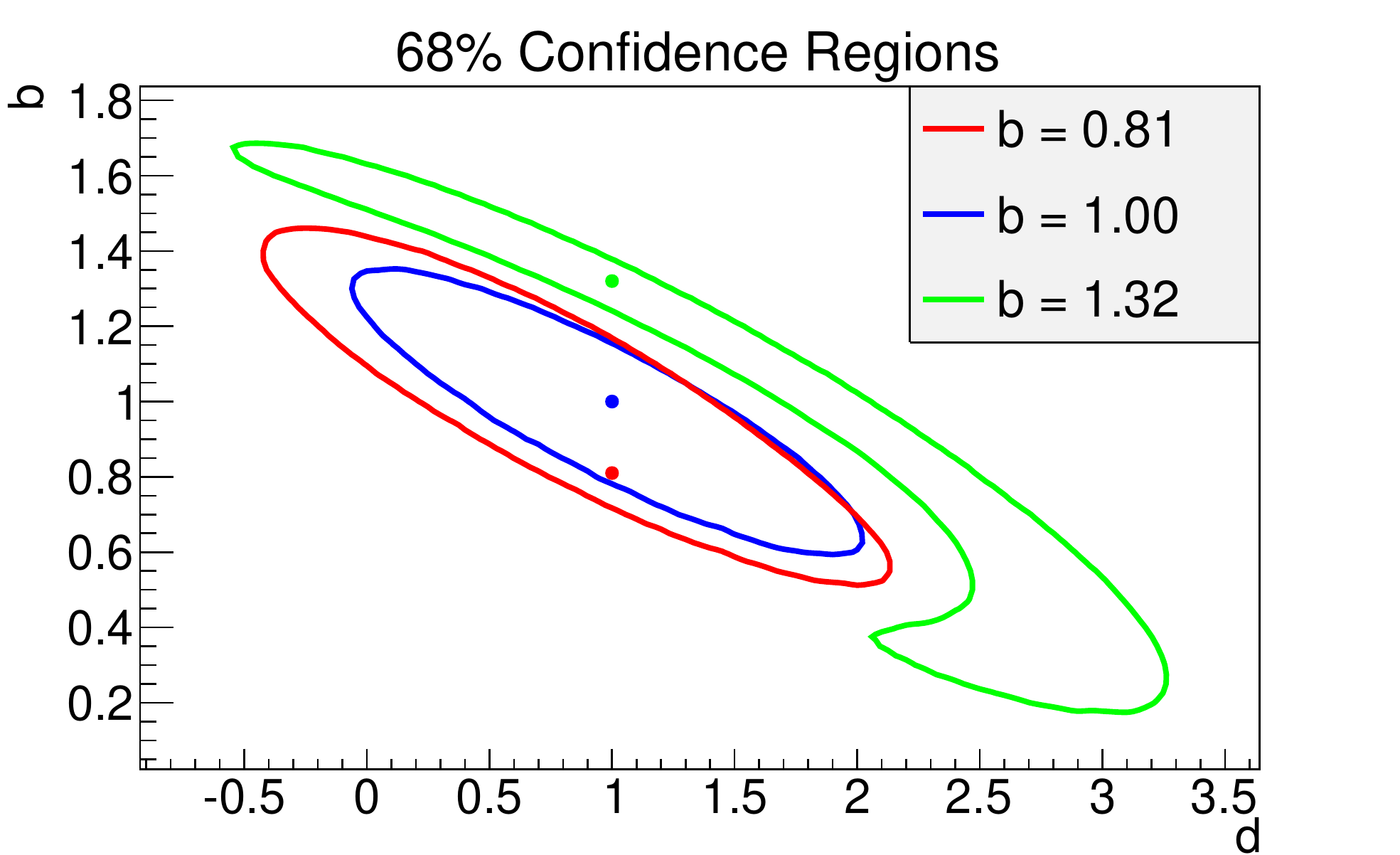}}
\resizebox{0.5\textwidth}{!}{\includegraphics{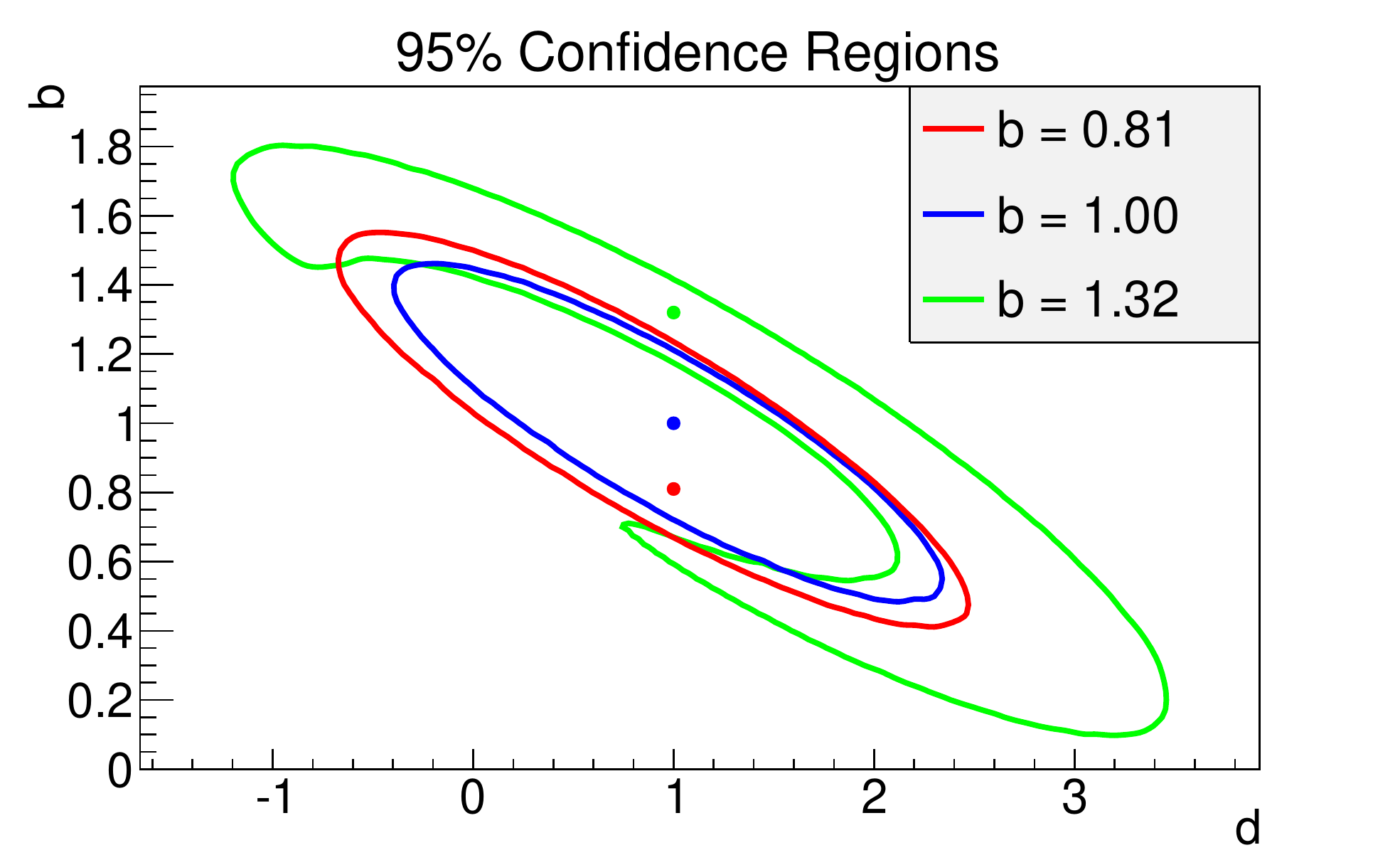}}
\caption{68\% and 95\% confidence regions ($\chi^2 = 2.28$ and 5.99, respectively) for the singlet ($b = 0.81$), doublet ($b = 1$) and GM ($b = 1.32$) models, with $d = 1$.  In all cases we fix $a = 0.9$.}
\label{fig:chisq}
\end{figure}

We plot the 68\% and 95\% confidence regions for our three benchmark points based on these two rate measurements in Fig.~\ref{fig:chisq}.  We find that the GM model with $a = 0.9$ can be distinguished from the doublet and singlet models at 68\% confidence level, and that the overlap at 95\% confidence level is small.  The doublet and singlet models cannot be distinguished using only these two event rate measurements.
The crescent shape of the 95\% confidence region for the GM model is caused by the WBF cross section not being monotonic in $b$, as can be seen from Table~\ref{tab:bvalues}.

Finally we note that the study in Ref.~\cite{Fujii-Princeton} assumes $M_h = 120$~GeV and both Higgs bosons decaying to $b \bar b$.  At 125~GeV, the Higgs decay branching fraction to $b \bar b$ is smaller, reducing the signal cross sections by about 20\%~\cite{Dittmaier:2012vm}.  The lost precision is expected to be recoverable by including $hh \to WWb \bar b$~\cite{Junping-LCWS}.

\subsection{$M_{hh}$ as a kinematic discriminant}

The ILD collaboration has developed a method to improve the sensitivity to the Higgs self-coupling $d$ by weighting events according to the invariant mass $M_{hh}$ of the Higgs pair~\cite{Junping-LCWS}.  This improves the precision on the extracted value of $d$ by about 10\% at both 500~GeV and 1~TeV in the SM case ($a = b = 1$)~\cite{Fujii-Princeton}.
This method could be adapted to improve the simultaneous sensitivity to $d$ and $b$ because the values of these couplings affect the shape of the $M_{hh}$ distribution as well as the total rate.  

\begin{figure}
\resizebox{0.5\textwidth}{!}{\includegraphics{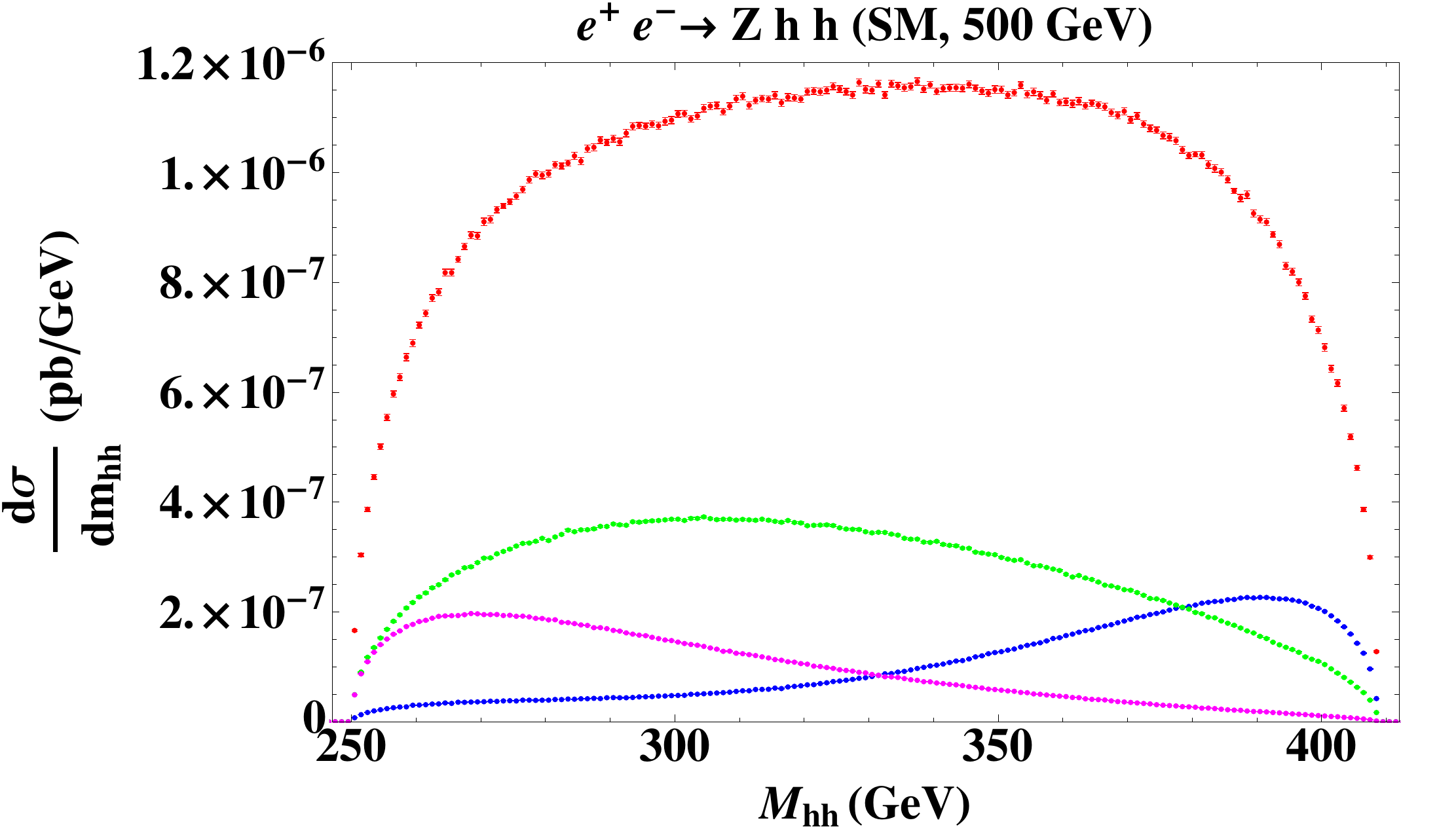}}
\caption{Differential cross section as a function of $M_{hh}$ for $e^+e^- \to Zhh$ in the SM at 500~GeV.  The four curves correspond to contributions from different sets of diagrams in Fig.~\ref{fig:eeZhh}: All diagrams (red uppermost points), diagram a (blue
or darkest line), diagram b (green or lightest line), and diagram c
(magenta or medium gray line)}
\label{fig:mhhzhh}
\end{figure}

In Fig.~\ref{fig:mhhzhh} we plot the SM cross section for $e^+e^- \to Zhh$ at 500~GeV as a function of $M_{hh}$, showing separately the contributions of the three diagrams in Fig.~\ref{fig:eeZhh} as well as the total cross section.  (Interference among the diagrams contributes significantly to the total cross section.)  The contribution to the amplitude from the diagram involving $d$ is largest at low $M_{hh}$, while the contribution from the diagram involving $b$ has a broader $M_{hh}$ distribution.

\begin{figure}
\resizebox{0.5\textwidth}{!}{\includegraphics{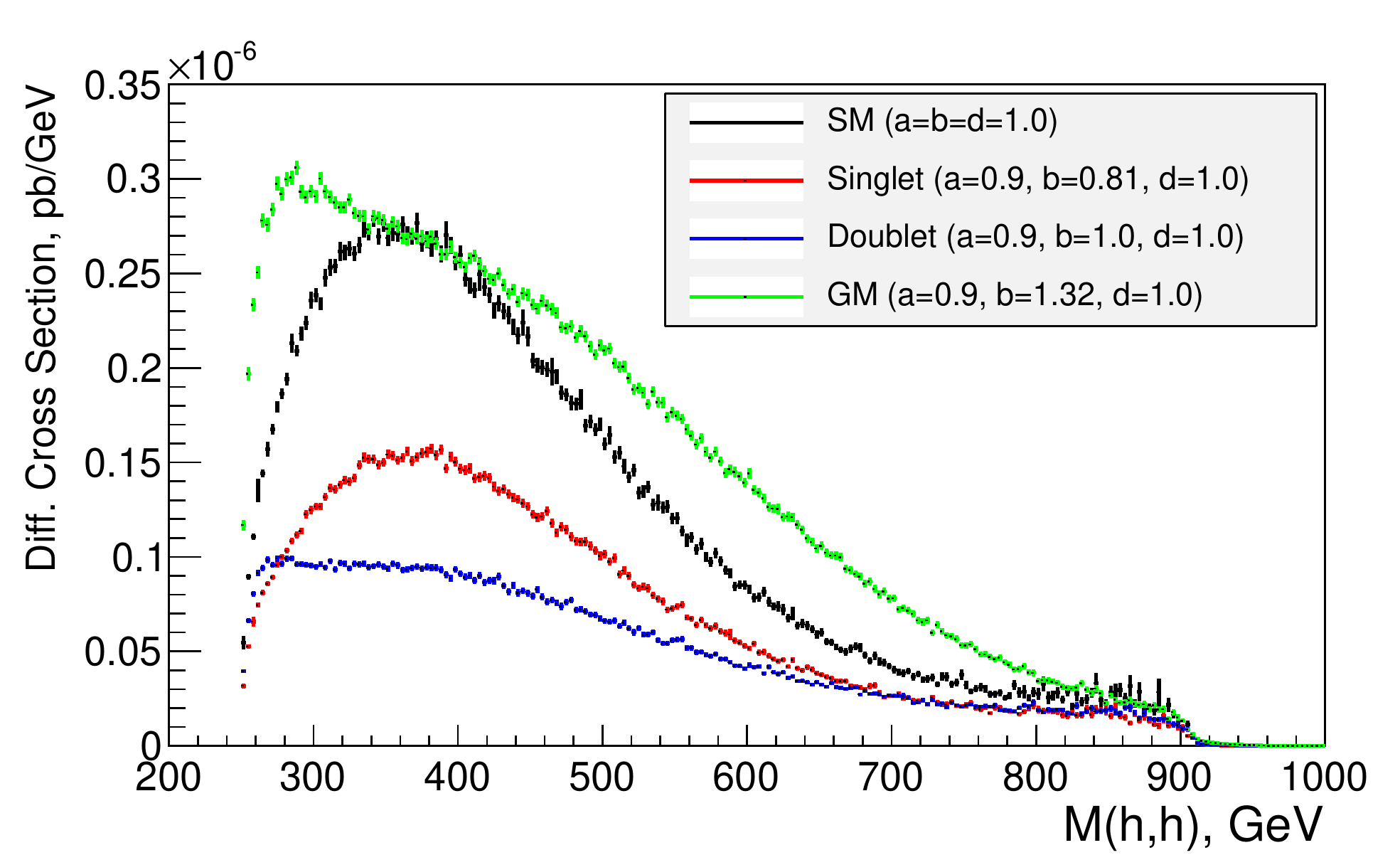}}
\caption{Differential cross section as a function of $M_{hh}$ for the WBF process $e^+e^- \to \nu \bar \nu hh$ at 1~TeV.  Shown are the distribution for the SM and our three benchmark points.}
\label{fig:mhhvvhh}
\end{figure}

In Fig.~\ref{fig:mhhvvhh} we plot the cross section for WBF $e^+e^- \to \nu \bar \nu hh$ at 1~TeV as a function of $M_{hh}$ for the SM and our three benchmark model points.  Notice in particular that our three benchmark models all have $a = 0.9$, $d = 1$, and differ only in their $b$ values.  The $b$ value has a dramatic effect on the $M_{hh}$ spectrum, due in part to interference among the three diagrams in Fig.~\ref{fig:eeWBFhh}.  Diagrams (b) and (c) interfere constructively, leading to the enhancement in the differential cross section at low $M_{hh}$ for higher $b$ values (compare the distributions for the GM and doublet models to that for the singlet model in Fig.~\ref{fig:mhhvvhh}).  Diagrams (a) and (b) interfere destructively, leading to the flattening of the spectrum at intermediate $M_{hh}$ for higher $b$ values (compare the distribution for the doublet model to that for the singlet model in Fig.~\ref{fig:mhhvvhh}).

A more sophisticated analysis taking into account the $M_{hh}$ distribution as well as the total cross section at each collision energy could thus provide additional sensitivity to $b$ and $d$.

\subsection{Synergy with LHC}

The Higgs self-coupling $d$ can also be accessed through double Higgs production in gluon fusion at the LHC~\cite{Baur:2002rb}.  In addition to $d$, the LHC double Higgs production cross section depends on the Higgs coupling to top quarks $c_t$.  It can also receive contributions from new colored particles that run in the gluon-fusion loop (new contributions to the $gg \to hh$ box diagram can be especially large)~\cite{Dawson:2012mk}, effective operators induced by such new particles~\cite{Pierce:2006dh}, or a direct $t \bar t hh$ coupling that can arise in composite-Higgs models~\cite{Manohar:2006gz}.  Nevertheless, double Higgs production at the LHC is insensitive to the $hhVV$ coupling, and can thus be used to constrain $d$ independently of this coupling.

A recent phenomenological analysis~\cite{Goertz:2013kp} found that $d$ could be constrained to be positive at 95\% confidence level using 600~fb$^{-1}$ at the 14~TeV LHC; with 3000~fb$^{-1}$ the $1\sigma$ uncertainties are reduced to $+30\%$ and $-20\%$.  The study assumed $c=1$ and no new particles in the loop.  A joint analysis of LHC and ILC double Higgs production would thus allow simultaneous constraints to be placed on $b$, $d$, and contributions from new colored particles or higher-dimensional operators.

\section{Conclusions}
\label{sec:conclusions}

Direct measurement of the $hhVV$ coupling will be of great interest if measurements of the couplings of the recently-discovered Higgs boson to SM particles reveal a deviation from the SM expectation.  The $hhVV$ coupling can be accessed through double Higgs production in $e^+e^-$ collisions.  In this paper we showed that separating the $hhVV$ coupling from the triple Higgs coupling can be accomplished using rate measurements at two different center-of-mass energies.  Additional sensitivity could be obtained by using the two Higgs invariant mass as a discriminant.  Furthermore, LHC measurements of double Higgs production in gluon fusion are insensitive to the $hhVV$ coupling and can be used to independently constrain the triple Higgs coupling.

Here we considered a simple set of benchmarks in which the new scalar(s) that mix with the SM Higgs preserve the custodial SU(2) symmetry.  This assumption is violated by a number of well-motivated models.  New scalar(s) that violate custodial SU(2) symmetry, such as the complex triplet in the type-2 seesaw mechanism for neutrino masses~\cite{seesaw}, yield $b_W \neq b_Z$.  In this case, two rate measurements are insufficient to simultaneously extract $b_W$, $b_Z$, and $d$, and additional information from kinematic discriminants and/or LHC measurements would be needed.

\begin{acknowledgments}
We thank K.~Hartling for collaboration on the Georgi-Machacek model and
T.~Gr\'egoire, P.~Kalyniak and K.~Moats for useful discussions.
K.K.\ and H.E.L.\ were supported by the Natural Sciences and Engineering Research Council of
Canada.
\end{acknowledgments}

\appendix
\section{Mixing with a vev-carrying scalar}
\label{app:vev}

Given a model-independent measurement of $a$, the extraction of $b$ and $d$ from double Higgs production measurements described in this paper does not depend on any assumptions about the vev of $X$.  Such an assumption enters only in the interpretation of the measurements of $a$ and $b$ in terms of the electroweak quantum numbers of $X$ when $X$ is not a doublet or singlet.  The vev dependence enters through the extraction of the $\phi$--$\chi$ mixing angle $\theta$ from the measurement of $a$.  Here we show that this mixing angle can still be extracted by taking advantage of the measurement of the fermion coupling $c$ in the case that the vev of $X$ is nonzero.  The assumption that $X$ does not couple to fermions is satisfied automatically when $X$ is not a doublet.

The real neutral state $\chi$ couples to $W$ pairs, and the vev of its parent multiplet contributes to the $W$ mass, via the Lagrangian term
\begin{equation}
	\mathcal{L} \supset \frac{g^2}{4} b_W^{\chi} (v_{\chi} + \chi)^2 W_{\mu}W^{\mu},
\end{equation}
where $b_W^{\chi}$ is given in Eq.~(\ref{eq:bchidef}), and similarly for the couplings and mass of the $Z$.  The $hVV$ couplings $a_V$ can then be written as
\begin{equation}
	a_V = \cos\theta \, v_{\phi} - b_V^{\chi} \sin\theta \, v_{\chi}.
\end{equation}
The mass of the $W$ imposes the additional constraint
\begin{equation}
	v_{\phi}^2 + b_W^{\chi} v_{\chi}^2 = v_{\rm SM}^2,
\end{equation}
and similarly for the $Z$ mass (we assume $\rho \simeq 1$).  From this we define
\begin{equation}
	\sin\beta = \frac{v_{\phi}}{v_{\rm SM}}, \qquad
	\cos\beta = \sqrt{1 - \frac{v_{\phi}^2}{v_{\rm SM}^2}} 
	= \sqrt{b_W^{\chi}} \frac{v_{\chi}}{v_{\rm SM}}.
\end{equation}
The $hVV$ couplings can then be re-expressed as
\begin{equation}
	a_V = \cos\theta \sin\beta - \sqrt{b_V^{\chi}} \sin\theta \cos\beta.
\end{equation}

Because an electroweak multiplet larger than a doublet cannot couple to charged fermions at tree level,  only the SM doublet $\phi$ contributes to fermion masses.  The coupling of the mass eigenstate $h$ to fermions is then given by
\begin{equation}
	c = \frac{\cos\theta}{v_{\phi}/v_{\rm SM}} = \frac{\cos\theta}{\sin\beta}.
\end{equation}
Note that when $v_{\chi} = 0$, we recover $c = a = \cos\theta$.  For nonzero $v_{\chi}$, we can solve for $\cos\theta$ in terms of the observables $a$ and $c$, given a model assumption for $b_V^{\chi}$.  The resulting expression for $\cos\theta$ can then be inserted into Eq.~(\ref{eq:bV}) for $b_V$, yielding a prediction for the chosen model.

\section{Constraints on our benchmark points}
\label{app:unitarity}

In our benchmark models, a full calculation of double Higgs production would include single production of the heavier custodial singlet with $H^0 \to hh$.  In the doublet and GM models, it would also include contributions from $t$- and $u$-channel exchange of the custodial SU(2) triplet states $(H^{\pm}, A^0)$.  We have neglected these contributions by computing double Higgs production cross sections using the effective theory of Eq.~(\ref{eq:efflagrangian}).  This is a good approximation when the additional Higgs particles are assumed to be heavy enough that on-shell processes such as $e^+e^- \to ZH^0$ with $H^0 \to hh$ and $e^+e^- \to A^0 h$ with $A^0 \to Zh$ are kinematically forbidden.  Here we discuss the viability of this assumption for our benchmark points.

An upper bound on the mass of the heavier custodial singlet $H^0 \equiv \phi \sin\theta + \chi \cos\theta$ can be obtained from the perturbative unitarity of $V_L V_L \to V_L V_L$ scattering (here $V_L$ denotes a longitudinally-polarized $W$ or $Z$ boson).  A coupled channel analysis including $ZZ$ and $WW$ in the initial and final states yields~\cite{unitarity}\footnote{We neglect the contribution from $t$-channel exchange of the custodial five-plet in the GM model~\cite{Falkowski:2012vh}.  The five-plet contribution is small for small $v_{\chi}$.}
\begin{equation}
	m_{H^0}^2 \lesssim \frac{16 \pi v_{\rm SM}^2}{5 (1 - a^2)} \simeq (1790~{\rm GeV})^2,
	\label{eq:H0bound}
\end{equation}
where we neglect terms of order $m_h^2$ and $m_W^2$ compared to $4 \pi v_{\rm SM}^2$ and take $a = 0.9$.
Perturbative unitarity constraints thus do not prevent us from assuming that resonant $H^0 \to hh$ contributions are beyond the kinematic reach of the 1~TeV ILC.  However, we note that if the $H^0hh$ coupling is large, processes involving off-shell $H^0 \to hh$ can have a significant effect on double Higgs production cross sections even for $H^0$ masses above the kinematic limit.  The $H^0$ contribution would be largest at the high end of the $M_{hh}$ distribution, which could potentially be used to discover the $H^0$ in this scenario.

Exchange of the custodial SU(2) triplet states $(H^{\pm}, A^0)$ present in the doublet and GM models is in fact required in order to restore perturbative unitarity of the $V_L V_L \to hh$ amplitude at high energies when $b - a^2 \neq 0$.  Including both $ZZ$ and $WW$ in the initial state, this requirement puts an upper bound on the custodial triplet masses (assumed degenerate) of~\cite{unitarity}
\begin{eqnarray}
	{\rm Doublet \! :} \ && m_{H^{\pm},A^0}^2 \lesssim \frac{8 \pi v_{\rm SM}^2}{\sqrt{3} (1 - a^2)}
		\simeq (2150~{\rm GeV})^2, \nonumber \\
	{\rm GM \! :} \ && m_{H^{\pm}_3,A_3^0}^2 \lesssim \frac{3 \pi v_{\rm SM}^2}{\sqrt{3} (1 - a^2)}
		\simeq (1320~{\rm GeV})^2,
\end{eqnarray}
where we again make the approximation $4 \pi v_{\rm SM}^2 \gg m_h^2, m_W^2$ and take $a = 0.9$.  Here the different coefficients for the doublet and GM models come from the different SU(2)$\times$U(1) quantum numbers of the custodial triplet states.

We also need to demonstrate that sufficient mixing can actually be obtained for each of our benchmark models.  We shall consider each scenario in turn.  We also comment on precision electroweak constraints.

The most general model that mixes the SM Higgs with a real neutral singlet contains dimensionful couplings that allow a sizeable $\phi$--$s$ mixing angle even when the singlet is heavy, without requiring dangerously large quartic scalar couplings.  The main constraint on the singlet model then comes from electroweak precision observables~\cite{Gupta:2012mi}, which take the simple
form
\begin{equation}
	S = \cos^2 \theta \, S_{\rm SM}(m_h) + \sin^2\theta \, S_{\rm SM}(m_{H^0}),
\end{equation}
where $S_{\rm SM}(m_i)$ is the SM contribution to the $S$ parameter evaluated for the SM Higgs mass equal to $m_i$,
and similarly for the $T$ parameter.  For most of the heavy scalar mass range we are interested in ($910 ~\text{GeV} \lesssim m_{H^0} \lesssim 1790 ~\text{GeV}$) the mixing angle is constrained by electroweak precision data to satisfy $\sin^2\theta \lesssim 0.1$ at the 90\% confidence level~\cite{Gupta:2012mi}. 
Our benchmark point corresponds to $\sin^2\theta = 0.19$, representing a mild violation of the precision electroweak constraints.  This can be compensated for with additional new physics that contributes to the $S$ and $T$ parameters.

In the doublet model it is harder to obtain a mixing angle as large as we have assumed while keeping $H^0$, $A^0$, and $H^{\pm}$ above their direct-production kinematic thresholds at the 1~TeV ILC.  We study this by scanning the parameter space of the CP-conserving two Higgs doublet model potential defined in Ref.~\cite{Gunion:2002zf} with a softly-broken $Z_2$ symmetry (corresponding to $m_{12}^2 \neq 0$ but $\lambda_6=\lambda_7=0$ in the notation of Ref.~\cite{Gunion:2002zf}) using the publicly available code {\tt 2HDMC}~\cite{Eriksson:2009ws}.  We find that obtaining $\cos\theta \equiv \sin(\beta-\alpha) = 0.9$ is possible under our mass constraints, but requires rather large quartic Higgs couplings (in particular, $\lambda_3$ and $\lambda_4$ of order 10).  This is a consequence of the decoupling property of the two Higgs doublet model~\cite{Gunion:2002zf}.  These large quartics lead to a large splitting between the $A^0$ and $H^{\pm}$ masses, which in turn leads to contributions to the $T$ parameter that push it outside the allowed experimental range.  As in the singlet model, this can be compensated with additional isospin-violating new physics.

In the GM model, the masses of the heavy scalars and the degree of mixing with the SM doublet can be independently controlled using two different dimensionful parameters, so that our benchmark conditions can be obtained without large quartic couplings.  To check this we scanned the parameters of the most general custodial SU(2)-preserving scalar potential~\cite{GMpaper},
\begin{eqnarray}
	V &=& \frac{\mu_2^2}{2} {\rm Tr}(\Phi^\dagger \Phi) 
	+  \frac{\mu_3^2}{2} {\rm Tr}(X^\dagger X) 
	 +  \lambda_1 [{\rm Tr}(\Phi^\dagger \Phi)]^2  \nonumber \\
         && + \lambda_2 {\rm Tr}(\Phi^\dagger \Phi) {\rm Tr}(X^\dagger X)   
         + \lambda_3 {\rm Tr}(X^\dagger X X^\dagger X) \nonumber \\
          &&   + \lambda_4 [{\rm Tr}(X^\dagger X)]^2   
                  - \lambda_5 {\rm Tr}( \Phi^\dagger \tau^a \Phi \tau^b) {\rm Tr}(X^\dagger t^a X t^b)       \nonumber \\
          && + M_1 {\rm Tr}(\Phi^\dagger \tau^a \Phi \tau^b)(X)_{ab} \nonumber \\
          && +  M_2 {\rm Tr}(X^\dagger t^a X t^b)(X)_{ab},
          \label{eq:GMpotential}
\end{eqnarray} 
where the doublet and triplet fields are written as
\begin{equation}
	\Phi = \left( \begin{array}{cc} \phi^{0*} & \phi^+ \\
	- \phi^{+*} & \phi^0 \end{array} \right), 
	\qquad
	X = \left( \begin{array}{ccc} \chi^{0*} & \xi^+ & \chi^{++} \\
	- \chi^{+*} & \xi^0 & \chi^+ \\
	\chi^{++*} & - \xi^{+*} & \chi^0 \end{array} \right).
\end{equation}
The potential in Eq.~(\ref{eq:GMpotential}) is identical to that studied in Ref.~\cite{Englert:2013zpa} except for the addition of the last two terms with coefficients $M_1$ and $M_2$, which are essential in order for the model to possess a phenomenologically-acceptable decoupling limit.  These two terms have traditionally been omitted for simplicity by imposing a discrete symmetry $X \to -X$ on the potential~\cite{Chanowitz:1985ug}.

We find that obtaining $\cos\theta = 0.9$ while keeping all additional states above their direct-production kinematic thresholds at the 1~TeV ILC can be achieved for large negative values of the dimensionful parameter $M_1$ (e.g., $M_1 \sim -2400$~GeV) and non-zero $v_\chi$ (e.g., $v_{\chi} \sim 30$~GeV) without requiring any large quartic scalar couplings.


\end{document}